\documentclass[aps, prb, twocolumn, superscriptaddress, citeautoscript, showpacs]{revtex4-2}
\usepackage{natbib}
\usepackage[dvips]{graphicx}
\usepackage[colorlinks, linkcolor = blue, citecolor = blue]{hyperref}
\usepackage{float}
\usepackage{graphics}
\setcitestyle{super}
\usepackage{bm}
\usepackage{changes}
\usepackage{amsmath}
\usepackage{array}

\begin{document}

\title{Long-range current-induced spin accumulation in chiral crystals}

\author{Arunesh Roy}
 \affiliation{Zernike Institute for Advanced Materials, University of Groningen, Nijenborgh 4, 9747AG Groningen, Netherlands}
\author{Frank T. Cerasoli}
\thanks{These three authors contributed equally.}
 \affiliation{Department of Physics, University of North Texas, Denton, TX 76203, USA}
\author{Anooja Jayaraj}
\thanks{These three authors contributed equally.}
  \affiliation{Department of Physics, University of North Texas, Denton, TX 76203, USA}
\author{Karma~Tenzin}
\thanks{These three authors contributed equally.}
   \affiliation{Zernike Institute for Advanced Materials, University of Groningen, Nijenborgh 4, 9747AG Groningen, Netherlands}
   \affiliation{Department of Physical Science, Sherubtse College, Royal University of Bhutan, 42007 Kanglung, Trashigang, Bhutan}
\author{Marco Buongiorno Nardelli}
 \affiliation{Department of Physics, University of North Texas, Denton, TX 76203, USA}
 \affiliation{The Santa Fe Institute, Santa Fe, NM 87501, USA}
\author{Jagoda S\l awi\'{n}ska}
\email{jagoda.slawinska@rug.nl}
 \affiliation{Zernike Institute for Advanced Materials, University of Groningen, Nijenborgh 4, 9747AG Groningen, Netherlands}
\date{\today}

\begin{abstract}
Chiral materials, similarly to human hands, have distinguishable right-handed and left-handed enantiomers which may behave differently in response to external stimuli. Here, we use for the first time an approach based on the density functional theory (DFT)+PAOFLOW calculations to quantitatively estimate the so-called collinear Rashba-Edelstein effect (REE) that generates spin accumulation parallel to charge current and can manifest as chirality-dependent charge-to-spin conversion in chiral crystals.  Importantly, we reveal that the  spin accumulation induced in the bulk by an electric current is intrinsically protected by the quasi-persistent spin helix arising from the crystal symmetries present in  chiral systems with the Weyl spin-orbit coupling. In contrast to conventional REE, the spin transport can be preserved over large distances, in agreement with the recent observations for some chiral materials. This allows, for example, generation of spin currents from spin accumulation, opening novel routes for the design of solid-state spintronics devices.
\end{abstract}

\maketitle
\section{Introduction}
Several phenomena in nature are governed by the geometric property called chirality. Chiral molecules and crystals have a property of handedness arising from the lack of inversion, mirror and roto-inversion symmetries, and host a plethora of intriguing effects manifesting differently in opposite enantiomers.\cite{chirality_biology, chirality_optics, chirality_plasmonics, chiral_light} Among these phenomena, the chirality-induced spin selectivity (CISS) describing the generation of a collinear spin current by a charge current flowing through a chiral molecule or assembly, is one of the most intriguing.\cite{adma_recent, review_apl}  A similar effect was recently realized in solid-state materials with strong spin-orbit coupling (SOC), opening a perspective for the design of devices based on the robust properties protected by crystal symmetries.\cite{tellurium_main, new_Te, nanowires, crnb3s6, yamamoto, disilicides} The observation of the chirality-dependent charge-to-spin conversion (CSC) was assigned to the collinear Rashba-Edelstein effect (REE) using the the symmetry analysis and models. However, the reliable quantitative description of spin accumulation induced by electric currents in chiral materials is still lacking.

Here, we develop and implement a computational approach to calculate current-induced spin accumulation in periodic systems based on density functional theory (DFT) and tight-binding Hamiltonians generated in the PAOFLOW code.\cite{paoflow1, paoflow2} We further perform calculations of the magnetization induced by electric currents in Te and TaSi$_2$, two different chiral materials with strong spin-orbit coupling (SOC) and compare them with experiments. Since the spin accumulation parallel to a charge current arises from the radial spin texture determined by the chiral point group symmetry,  we analyze the landscape and symmetry of the spin-orbit field to explain in detail the properties of spin transport in chiral crystals. In particular, we demonstrate that the Weyl-type SOC term locally describing the radial spin texture, may generate a quasi-persistent spin helix in the real space which prevents the spin randomization and results in a very long spin lifetime in a diffusive transport regime.  The collinear spin accumulation induced in chiral materials, in contrast to the conventional REE, is thus intrinsically protected from the spin decoherence which may have a far-reaching impact on the design of spintronics devices.

\section{Results and Discussion}
\subsection{Calculation of the current-induced spin accumulation for arbitrary SOC}
The spin accumulation also referred to as current-induced spin polarization (CISP), is a non-equilibrium magnetization induced by an electric current that flows through a non-magnetic material with a large SOC.\cite{ganichev, ivchenko, aronov1, aronov2} Even though it was discovered decades ago, experimental studies typically addressed its special case, the Rashba-Edelstein effect present in systems with the Rashba-type spin-momentum locking, such as a two-dimensional electron gas (2DEG), interfaces and surfaces with broken inversion symmetry.\cite{johansson_sto, johansson_alge, ghiasi} The fact that the spin accumulation might arise from any other type of SOC, such as Dresselhaus, Weyl, or more complex spin arrangements, was often disregarded and alternative configurations were less studied, being mostly limited to models.\cite{tsymbal_cisp}

\begin{figure*}
\includegraphics[scale = 1.0]{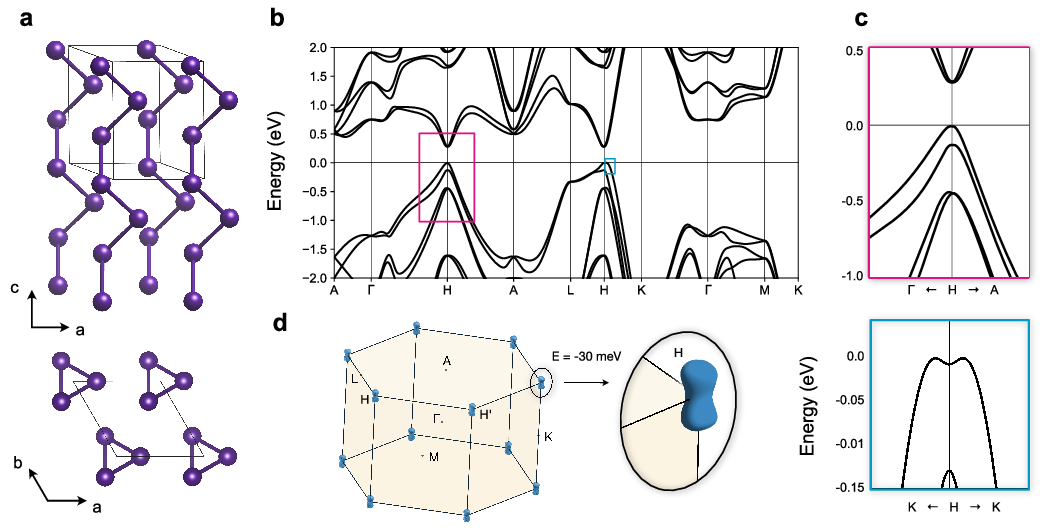}
\caption{
\label{struct}
Crystal structure and electronic properties of chiral tellurium. (a) Side and top view of the left-handed Te (SG 154).
(b) Band structure calculated along the high-symmetry lines in the Brillouin zone (BZ), identical for both enantiomers. (c) Zoom on the two regions marked by the color rectangles in (b). The upper panel shows the set of $5p$ lone-pair bands around the H point. The bottom panel illustrates the details of the topmost valence band. (d) Brillouin zone with the high-symmetry points and Fermi surface at $E=-30$ meV; the inset shows a zoom-in of the dumbbell-shaped hole pocket at the H point.
}
\end{figure*}

\begin{figure*}
\includegraphics[scale = 1.0]{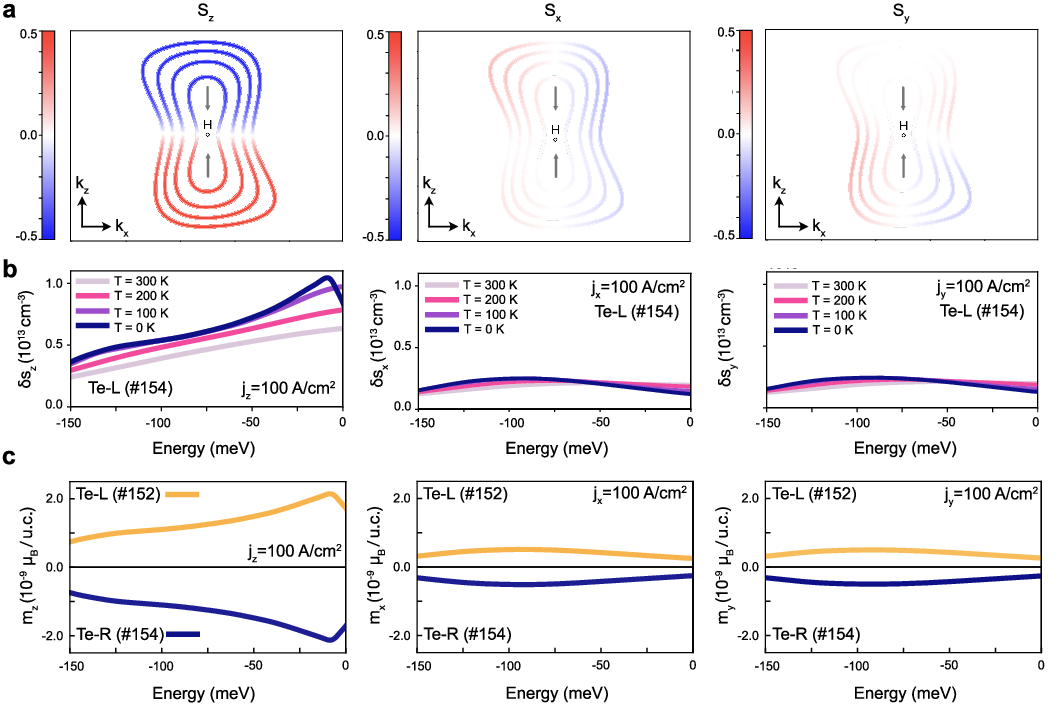}
\caption{
\label{fig:spin_bs_cisp}
Spin-resolved electronic structure and spin accumulation in bulk Te. (a) The calculated isoenergy contours projected onto the $k_x-k_z$ plane passing through the H point in the left-handed Te. The energy eigenvalues are selected at 10~meV, 30~meV, 50~meV, and 70~meV below the Fermi level, counting from the center. The color maps superimposed on the contours represent the expectation values of $S_x$, $S_y$ and $S_z$ shown at the left, middle and right panel, respectively.  The spin texture is radial with respect to the H point and nearly persistent along $k_z$, as indicated by the arrows. (b) Spin accumulation per volume induced by the charge current $j=100$ A\,cm$^{-2}$ applied along $z$, $x$ and $y$ direction in the left-handed Te, represented by left, middle, and right panel, respectively. Different lines correspond to temperatures in the range from 0 K to 300 K. (c) The corresponding magnetization per unit cell calculated for the left-handed (blue lines) and right-handed Te (orange lines) at 0 K.
}
\end{figure*}

We will start with a brief overview of the implemented computational approach that allows the calculation of the spin accumulation for arbitrary SOC. The equilibrium electron distribution in a crystal can be described by the Fermi distribution function $f^{0}_{\mathbf{k}}$. Due to the time reversal symmetry, the expectation values of the spin operator $\mathbf{S}$ at the opposite momenta cancel, resulting in a zero net spin polarization. Upon the flow of charge current, the spin polarized Fermi surface shifts as a result of the applied field and the non-equilibrium distribution $f_{\mathbf{k}} = f^{0}_{\mathbf{k}}+\delta f_{\mathbf{k}}$ generates a net spin polarization:
\begin{equation}
\delta \mathbf{s} = \sum_{\mathbf{k}} \langle \mathbf{S} \rangle_{\mathbf{k}} \delta f_{\mathbf{k}}
\end{equation}
We can calculate $\delta f_{\mathbf{k}}$ in the framework of the semi-classical Boltzmann transport theory and write:


\begin{equation}
\label{delta_s}
\delta \mathbf{s} = \sum_{\mathbf{k}} \langle\mathbf{S}\rangle_{\mathbf{k}}\tau_{\mathbf{k}} \left(\mathbf{v}_{\mathbf{k}} \cdot \mathbf{E}\right) \frac{\partial f_{\mathbf{k}}}{\partial E_{\mathbf{k}}}
\end{equation}
where $\tau_{\mathbf{k} }$ is the relaxation time and $\mathbf{v}_{\mathbf{k}}$ the group velocity.

\noindent Because the charge current density can be expressed using the same $\delta f_{\mathbf{k}}$,\cite{ziman}
\begin{equation}
\label{current}
\mathbf{j_c} = -\frac{e}{V}\sum_{\mathbf{k}} \mathbf{v}_{\mathbf{k}} \delta f_{\mathbf{k}}.
\end{equation}
\noindent we can introduce the spin accumulation tensor $\chi$ defined as the ratio of the quantities from Eqs. (\ref{delta_s}) and (\ref{current}),
\begin{equation}
  \label{tensor}
  \chi^{ji} = -\cfrac{\sum\limits_{\mathbf{k}} \langle\mathbf{S}\rangle_{\mathbf{k}}^j\mathbf{v}_{\mathbf{k}}^i  \frac{\partial f_{\mathbf{k}}}{\partial E_{\mathbf{k}}}}{e \sum\limits_{\mathbf{k}} (\mathbf{v}_{\mathbf{k}}^i)^2 \frac{\partial f_{\mathbf{k}}}{\partial E_{\mathbf{k}}}}
\end{equation}
in which we assumed the constant relaxation time approximation. The induced spin accumulation per unit volume can be then calculated from the formula,
\begin{equation}
\label{cisp_main}
\delta s^j=\chi^{ji} j_{i}^\mathrm{A}
\end{equation}
where $j^A_i$ is the value of the charge current applied along an arbitrary $i$ direction. Consequently, we can calculate the induced magnetization per unit cell as,
\begin{equation}
  \label{mag}
  \mathbf{m}=-g_{s} \mu_{\mathrm{B}} V \delta \mathbf{s}/\hbar
\end{equation}
where $g_{s} = 2$ is the Land\'e g-factor, $\mu_{\mathrm{B}}$ is Bohr magneton and $V$ denotes the volume of the unit cell.

The quantities required to compute the tensor $\chi$ for any material can be evaluated using accurate \textit{ab~initio} tight-binding (TB) Hamiltonians constructed from self-consistent quantum-mechanical wavefunctions projected onto a set of atomic orbitals, as detailed in Methods.\cite{agapito0} Alternative approaches for the calculations of spin accumulation can be found in the previous studies.\cite{mokrousov, salemi}

\subsection{Unconventional charge-to-spin conversion in Te}
One of the simplest chiral crystals is trigonal tellurium that crystallizes in two enantiomorphic structures sharing the point group $D_3$, the right-handed described by the space group \textit{P3$_121$} (SG 152) and the left-handed belonging to the space group $P3_221$ (SG 154); the structure of the latter is schematically shown in Fig. 1a. Te atoms form covalently bonded spiral chains which are arranged hexagonally and interact with each other via weak van der Waals forces. The threefold screw symmetry $C_3$ that determines the chirality runs along the $c$ axis, while additional twofold rotational symmetry axes $C_2$ lie within the $a-b$ plane. The calculated electronic structure is presented in Fig.~1b-c and agrees well with the earlier theoretical and experimental findings.\cite{old1, old_farbshtein, agapito_te, souza, pressure, kuniko} While it is difficult to reproduce the semiconducting behavior of Te by standard simulations based on the generalized-gradient or local-density approximation, the calculations within the novel pseudo-functional approach ACBN0 yield a gap of 279 meV which is close to the measured value of 330 meV,\cite{gap_exp} and capture the characteristic 'camelback' shape of the topmost valence band (VB) with a local maximum along the H-K line.\cite{vanderbilt}

The strong spin-orbit interaction in Te combined with the lack of the inversion symmetry results in a large spin-splitting of the bands in the whole energy range, but only the topmost VB contributes to electronic and spin transport. Due to the intrinsic p-type doping coming from the vacancies, the Fermi level is shifted by few tens meV below the VBM and the Fermi surface consists of two small pockets around the H and H' points (Fig. 1d), with size and shape depending on the doping. The constant energy contours shown in Fig. 2a correspond to pockets of different sizes projected onto the $k_x-k_z$ plane passing through the H point, while the superimposed color maps represent $S_x$, $S_y$, and $S_z$ components of the spin texture.

The peculiar orientation of spins, radial toward the H and H' points is enforced by the three-fold screw symmetry along the K-H line and the twofold symmetry with respect to the H–H' lines in the absence of any mirror planes.\cite{pressure, tellurium_main} Such a spin arrangement, resembling a magnetic monopole in the reciprocal space, is a signature of the prototypical Weyl-type SOC, and may occur only in the systems lacking the mirror symmetry. Moreover, the $S_z$ component prevailing over $S_x$ and $S_y$, indicates that the spin texture is almost entirely aligned with the $k_z$, and suggests its quasi-persistent character. The radial spin texture of trigonal Te, including spin polarization nearly parallel to the long direction of the hole pocket, was confirmed by the recent spin- and angle-resolved photoemission spectrocopy (S-ARPES) measurements.\cite{vobornik, sakano}

Based on the accurate spin-resolved electronic structure, we calculated the chirality-dependent spin accumulation induced by an applied electric current. Because the spins around the H and H' points are parallel to the momentum, any projection of $\delta \mathbf{s}$ can be achieved as long as the charge current flows in the same direction. This is in line with the symmetry analysis that we performed in the spirit of Seemann \textit{et al.}\cite{wimmer2015spin, seemann2015symmetry, arunesh} We found that the $\chi$ tensor will indeed have only diagonal elements. The components $\delta s_{z}$,  $\delta s_{x}$ and $\delta s_{y}$ induced by the electric current of 100 A cm$^{-2}$ along the $z$, $x$ and $y$ directions, respectively, are plotted in the left, middle, and right panel of Fig 2b. Although the spin accumulation perpendicular to the screw axis was never reported for chiral Te, it is evident that the components $\delta s_{x}$ and $\delta s_{y}$, albeit lower than $\delta s_{z}$ by an order of magnitude, are present and could be observed in experiments.

The dependence of $\delta s_{z}$ on chemical potential (Fig.~2b), closely reflects the electronic structure and the spin texture of the topmost VB. The highest (negative) value of the spin accumulation coincides with the local band minimum at the H point (-8~meV, bottom panel in Fig.~1c), and it decreases when approaching the next valence band with the opposite spin polarization. Although reaching its edge at -130~meV with respect to $E_\mathrm{F}$ would require unrealistic values of doping (approximately $2.3\times10^{19}$ vs. $10^{14}-10^{17}$~cm$^{-3}$ typically observed in Te samples\cite{pnas1, pnas2}), the decrease in the magnitude can be noticed already above this value. Thus, the low concentrations of holes seem to maximize the spin accumulation in Te. Another factor that may play a relevant role is the temperature (see Fig.~2b). In particular, $\delta s_{z}$ tends to drop by a factor of two at room temperature, which cannot be easily explained and requires further systematic studies. In contrast, the components $\delta s_x$ and $\delta s_y$ do not strongly depend neither on doping nor temperature.
\vskip 1.0 em
Finally, we will compare the current-induced magnetization calculated from the Eq. (\ref{mag}) and displayed in Fig. 2c with the values obtained based on the nuclear magnetic resonance measurements.\cite{tellurium_main} Our predicted magnetization $10^{-9}\mu_\mathrm{B}$ per unit cell agrees well with the model conceived by Furukawa \textit{et al.},\cite{tellurium_main} but it is one order of magnitude lower than the value estimated from their experimental data (10$^{-8}\mu_\mathrm{B}$). In addition, our calculated dependence of the induced spin accumulation on chemical potential agrees very well with the experimental trend revealed recently for Te nanowires.\cite{nanowires}
Importantly, the collinear REE being directly linked to the spin texture, yields exactly opposite signs of $\delta \mathbf{m}$ for different enantiomers which are connected by the inversion symmetry operation (see Fig.~2c), which is in line with the observed chirality-dependent charge-to-spin conversion.


\begin{figure*}
\includegraphics[scale = 1.0]{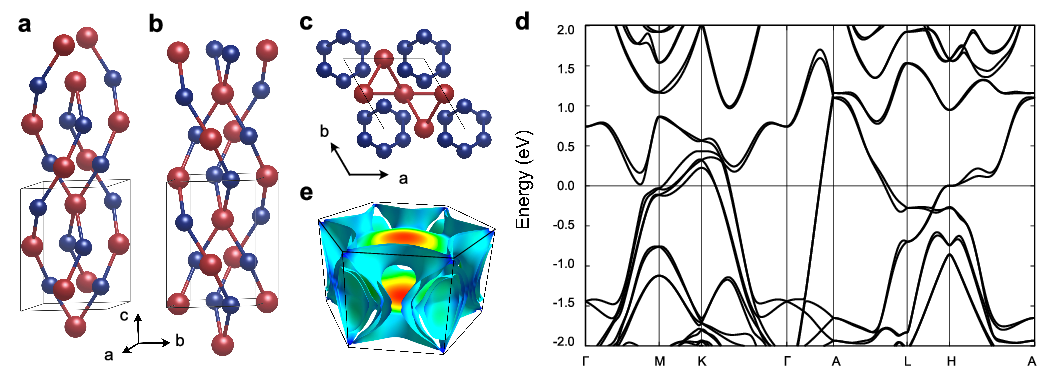}
\caption{
\label{fig:tasi2}
Structure and electronic properties of chiral TaSi$_2$. (a) Side view of the right-handed TaSi$_2$ (SG 180). Ta atoms are denoted as red and Si atoms as blue balls. (b) Top and side view of the left-handed TaSi$_2$ (SG 181). (c) The calculated Fermi surface ($E=E_F$) consisting of four nested sheets. The color scheme reflects the Fermi velocity.\cite{fermisurfer} (d) Band structure calculated along the high-symmetry lines in the momentum space. The high-symmetry points are consistent with the labels in Fig. 1.
}
\end{figure*}

\begin{figure*}
\includegraphics[scale = 1.0]{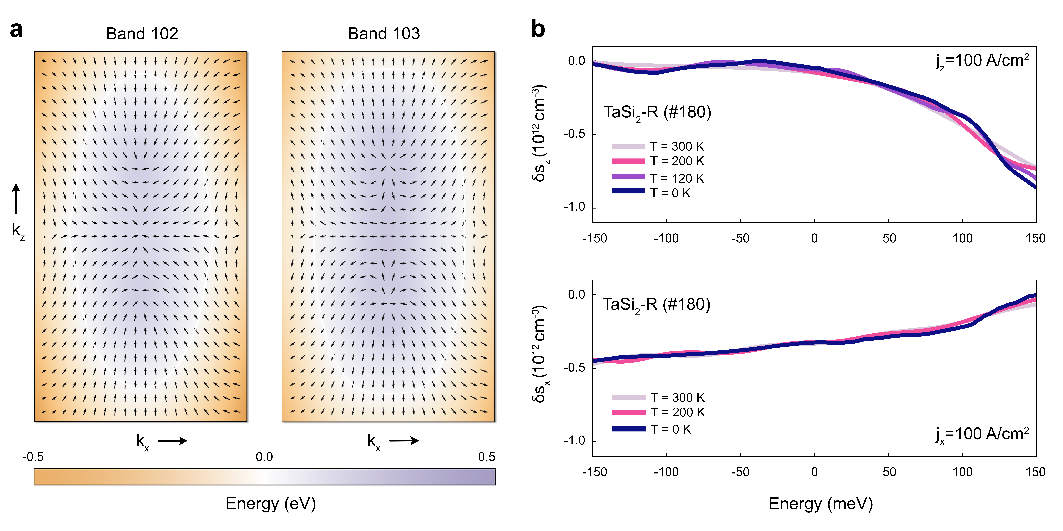}
\caption{
\label{fig:tasi2_results}
Spin texture and spin accumulation in the right-handed TaSi$_2$ (SG 180). (a) Energy eigenvalues and spin texture of the two innermost band visible in Fig. 3e projected on the $k_x-k_z$ plane passing through the $K$ point; the $K$ point is located exactly at the center. The $S_y$ components are included only in the norms of the spin vectors. The multiple arrows serve for a better illustration of the trend, but only those coinciding with the Fermi contour – denoted as the white line - contribute to the spin accumulation. (b) The calculated spin accumulation vs energy; the components $\delta s_{z}$ and $\delta s_x$ are shown in the upper and bottom panel, respectively. The omitted $\delta s_y$ is similar to $\delta s_x$. The colors of the lines correspond to different temperatures.
}
\end{figure*}

\subsection{CISP in a semimetallic disilicide TaSi$_2$.}
Inspired by the spin transport experiments that revealed signatures of chirality-dependent response in  disilicides, we performed the calculations for chiral TaSi$_2$.\cite{disilicides} Its right-handed structure belongs to the space group P6$_2$22 (SG 180) and the left-handed one is characterized by the space group P6$_4$22 (SG 181); both enantiomers are displayed in Fig. \ref{fig:tasi2}a-b. The hexagonal unit cell contains two pairs of intertwined Ta-Si-Ta chains with different helicity running along the $c$ direction, which is reversed for the opposite enantiomers. The left-handed and right-handed structures are distinguished by the sixfold screw symmetry $C_6$ along the $c$-axis, while the additional $C_2$ rotational symmetries are defined with respect to the crystal $a-b$ face diagonals (see Fig. \ref{fig:tasi2}c). The calculated electronic structure shown in Fig. 3d is in a good agreement with the existing theoretical and experimental reports.\cite{tasi2_japanese, degruyter, anisotropic} TaSi$_2$ is a Weyl semimetal and indeed, the energy dispersion contains several degenerate crossings protected by the nonsymmorphic symmetry, for example the nodes at the M and H points that lie close to the Fermi level. The shape of the Fermi surface (Fig. 3e), consisting of four non-intersecting sheets is also in line with the results of the previous de Haas-van Alphen experiments.\cite{tasi2_japanese}

The current-induced spin accumulation in the semimetallic TaSi$_2$ is governed by the spin texture of the Fermi surface. Because it contains multiple nested sheets with different spin patterns, it is not straightforward to analyze the full landscape of the spin-orbit field inside the BZ. Nevertheless, the linear-in-$k$ spin arrangement at the high-symmetry points can be predicted with the help of the group theory and further verified by the DFT calculations. Following the study by Mera Acosta \textit{et al.},\cite{zunger} we note that the crystallographic point group $D_6$ describing TaSi$_2$ may yield the purely radial Weyl spin texture at the $k$-vectors possessing the point group symmetry $D_3$ or $D_6$. Using the Bilbao Crystallographic Server,\cite{bcs1, bcs2} we determined little point groups of all the high-symmetry points ($\Gamma$, A, K, H and M) and we found that most of them can be the 'source' or 'sink' of the radial spin texture. Only the M point described by the little point group $D_2$ allows both Weyl and Dresselhaus spin arrangement.\cite{zunger} As an example, Fig. 4a illustrates the spin texture of the two innermost bands contributing to the Fermi surface, projected onto the $k_x-k_z$ plane around the K point. The spin textures are almost perfectly radial with a visible persistent component $S_z$ that seems to be enforced by the elongated band shapes. In a 3D $k$-space, these FS would resemble ellipsoids, one inside another, with the hedgehog-like spin textures.

The radial spin arrangements dominating in the BZ, generate the unconventional spin accumulation that may manifest as the spin selectivity in TaSi$_2$. The calculated $\delta \mathbf{s}$ induced by the electric current of 100 A/cm$^{2}$ flowing in the directions parallel and perpendicular to the screw axis is shown in Fig.~4b. The results are presented as the energy dependence to better show the connection with the band structure, but TaSi$_2$ is semimetallic and mostly the values close to the Fermi level are relevant for the transport. Surprisingly, $\delta s_z$ at $E_\mathrm{F}$ is nearly two orders of magnitude lower than in the slightly doped Te (10$^{11}$ cm$^{-3}$ vs 10$^{13}$ cm$^{-3}$), while $\delta s_x$ is lower by one order of magnitude. Such a difference seems to be due to the fact that two pairs of bands with the opposite polarization simultaneously cross the Fermi level. This contrasts with the case of Te which FS consists of the two identical spin-polarized pockets. In fact, the increase of the $\delta s_z$ magnitude above $E_\mathrm{F}$ in TaSi$_2$ can be assigned to the emergence of an isolated strongly spin-polarized pocket at +20 meV.

Chirality-dependent response in disilicides was suggested based on the spin transport experiments for left- and right-handed NbSi$_2$ and TaSi$_2$.\cite{disilicides} Although these reports were quantitative and cannot be compared with the result of the DFT calculations, we believe that the current-induced spin accumulation may give rise to the observed interconversion between charge and spin current. First, we note that the unconventional SHE in TaSi$_2$ (SG 180) is forbidden by the symmetry and could not contribute to the collinear spin transport.\cite{arunesh} Second, the quasi-persistent character of the spin texture enforced by the crystal symmetry (Fig. 4a) indicates that the spins accumulated in the bulk should be partially protected from scattering, and capable to diffuse to the detection electrode. Below, we will explicitly show that the anti-symmetric SOC arising from the monoaxial screw symmetry, and approximated by the linear-in-$k$ Weyl term in the Hamiltonian,\cite{zunger, hasanchiral} generates the \textit{near persistent spin helix} in the real space and gives rise to a potentially infinite spin lifetime in chiral materials.

\subsection{Protection of spin transport over macroscopic distances}
To complete the description of spin transport in chiral materials, we will demonstrate
the emergence of the \textit{persistent spin helix} that protects from dephasing the spin accumulation generated in the bulk. Let us consider the free electron dispersion $H_{0} = \hbar^2 \mathbf{k}^2/2m$ where $m$ is the effective electron mass and $\mathbf{k} = (k_x,k_y,k_z)$ and the Weyl spin-orbit coupling term $H_{so} = \alpha_x k_x \hat{\sigma}_x + \alpha_y k_y \hat{\sigma}_y + \alpha_z k_z \hat{\sigma}_z$, where $\hat{\sigma}_{i}$ denotes the Pauli spin matrices and $\alpha_{i}$ are the SOC parameters along $ i = x,y,z$ directions. Our first principles calculations show that the spin textures of Te and TaSi$_2$ are strongly enhanced along the direction of the screw axis, as illustrated in Fig. \ref{fig:spin_bs_cisp}a and Fig. \ref{fig:tasi2_results}a. More specifically, in the case of Te the SOC parameters are $\alpha_z >> \alpha_j$ for $j = x,y$, whereas in TaSi$_2$ we have $\alpha_z>\alpha_x>\alpha_y$. The Hamiltonian is then, without the loss of generality,
\begin{equation}\label{hamiltonian}
H = H_0 + H_{so} \simeq \frac{\hbar^2 k^2_{\parallel}}{2m} +  \frac{\hbar^2 k_{z}^2}{2m} + \alpha_{z} k_{z} \hat{\sigma}_{z},
\end{equation}
where $k^2_{\parallel} = k_x^2 + k_y^2$. The Hamiltonian is diagonal in the spin space and its eigenvalues are
\begin{equation}\label{spin_band}
E_{\uparrow, \downarrow} (\mathbf{k}) = \frac{\hbar^2 k^2_{\parallel}}{2m} +
\frac{\hbar^2}{2m}\left(k_z \mp \frac{m\alpha_z}{\hbar^2}\right)^2 - \frac{\hbar^2}{2m}\left(\frac{m\alpha_z}{\hbar^2}\right)^2,
\end{equation}
where $E_{\uparrow, \downarrow}$ denote bands with $\uparrow, \downarrow$ spin projections, respectively. The Eq.(\ref{spin_band}) satisfies the shifting property
\begin{equation}
E_{\uparrow}(\mathbf{k}) = E_{\downarrow}(\mathbf{k} + \mathbf{Q}),
\end{equation}
proposed by Berenevig \textit{et al.}, which implies that the Fermi surfaces consists of two circles, shifted by magic-shifting vector $\mathbf{Q} = 2 m\alpha_z/\hbar^2$.\cite{berenevigpst} Following the analysis similar to the one for the Rashba-Dresselhaus model with equal SOC parameters,\cite{berenevigpst} we identify the existence of the $SU(2)$ symmetry of the Hamiltonian in Eq.(\ref{hamiltonian}), which is robust against disorder and Coulomb interaction and can give rise to, ideally, infinite spin lifetime along the $z$-direction (see Fig. 5). The presence of the weak SOC along the $x$ and $y$ directions, perturbs the exact $SU(2)$ symmetry.  The \textit{near persistent spin helix} yields the finite, but sizable spin relaxation length and makes possible the detection of the spin accumulation induced by the electric current in the bulk chiral crystals.\cite{crnb3s6, disilicides}

\begin{figure*}
\includegraphics[scale = 1.0]{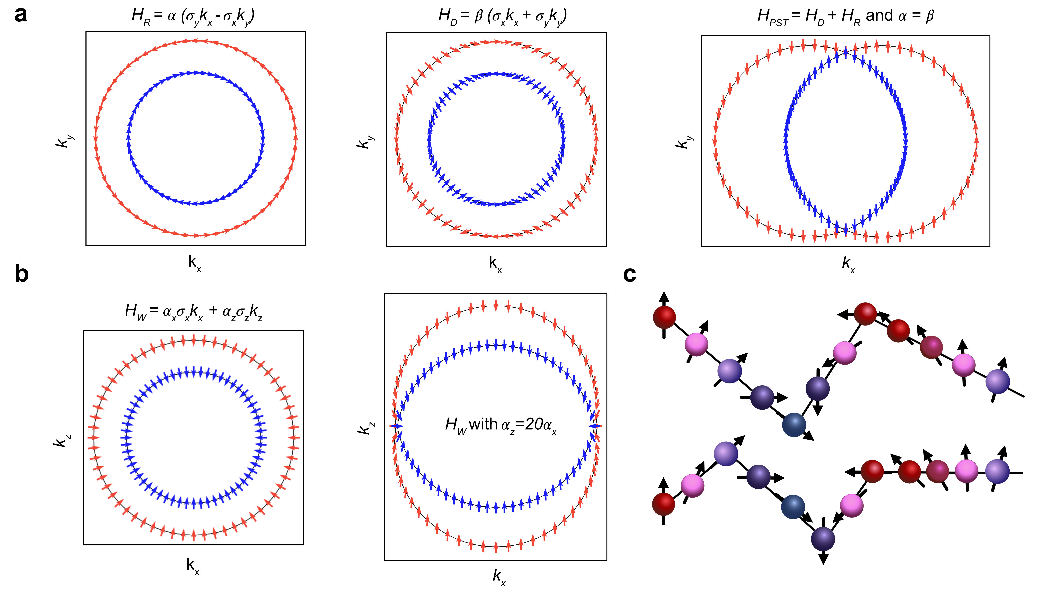}
\caption{
\label{fig:scheme}
Persistent spin texture and spin helix in the Rashba-Dresselhaus (RD) and Weyl (W) models enhancing spin transport in systems with strong SOC. (a) Pure Rashba (left) and Dresselhaus (middle) models; their combination with $\alpha=\beta$ (right) generates a persistent spin texture in the momentum space. (b) Analogous mechanism contributing to a persistent spin texture in the Weyl model describing the radial spin texture; Weyl Hamiltonian with equal ($\alpha_z=\alpha_x$) parameters (left) and with the prevailing $\alpha_z \sigma_z$ term (right). The latter, with $\alpha_z=20\alpha_x$ closely resembles the spin texture of Te (see Fig. 2a). (c) Schematic illustration of a persistent spin helix in the real space observed in quantum wells with balanced Rashba and Dresselhaus terms.\cite{koralek} The collective spin is a conserved quantity due to the emergence of the $SU(2)$ symmetry. The quasi-persistent spin texture described by the Weyl Hamiltonian will lead to a similar effect protecting the spin transport.
}
\end{figure*}

In summary, we calculated the collinear Rashba-Edelstein effect originating from radial spin textures, using for the first time the accurate DFT-based approach that allowed us estimate the induced magnetic moment per unit cell for two representative chiral materials Te and TaSi$_2$. While the values and dependence on chemical potential calculated for the former are in agreement with the recent experiments, confirming that the observed chirality-dependent charge-to-spin conversion indeed originates from spin accumulation,\cite{tellurium_main, nanowires} our approach implemented in PAOFLOW will allow for further calculations of spin transport in materials studied experimentally. Importantly, we also proved via the analysis of the calculated spin textures and a general Weyl Hamiltonian, that the chiral materials can host the quasi-persistent spin helix emerging in the real space as a consequence of the crystal symmetry. It protects spins against scattering and ensures a sizable spin relaxation length, which seems to be in line with the recent observation of micrometer-range spin transport. This makes the collinear REE different from the conventional one, where the bulk spin accumulation gets easily randomized. The possibility of electric generation of spin accumulation, preserving at the same time the long spin lifetime, will have important implications for spintronics devices.

\section{Methods}
\textbf{First-principles calculations}
We performed calculations based on the density functional theory (DFT) using the \textsc{Quantum Espresso} package.\cite{qe1,qe2} We treated the ion-electron interaction with the fully relativistic pseudopotentials from the pslibrary (0.2) database,\cite{pslibrary} and expanded the electron wave functions in a plane wave basis set with the cutoff of 80 Ry. The exchange and correlation interaction was taken into account within the generalized gradient approximation (GGA) parameterized by the Perdew, Burke, and Ernzerhof (PBE) functional.\cite{pbe} The Te crystals were modeled using hexagonal unit cells containing three atoms, which were fully optimized with the convergence criteria for energy and forces set to $10^{-5}$ Ry and $10^{-4}$ Ry/bohr, respectively. We applied the Hubbard correction calculated self-consistently using the ACBN0 method,\cite{acbn0} which value $U_\textrm{5p}$ = 3.81 eV was included in the relaxations and the electronic structure calculations. The lattice constants were optimized to $a=b=4.51$ \AA\, and $c=5.86$ \AA. TaSi$_2$ was modeled in a hexagonal unit cell containing nine atoms. The lattice parameters were fixed to the experimental values $a=b=4.78$ \AA\, and $c=6.57$ \AA,\cite{degruyter} while the internal coordinates were relaxed. The BZ integrations were performed using the Monkhorst-Pack scheme with $k$-points grids of $22\times 22\times 16$ and $16\times 16\times 12$ for Te and TaSi$_2$, respectively. The Gaussian smearing of 0.001 Ry was chosen as the orbital occupation scheme. SOC was included self-consistently in all the calculations.

\textbf{Spin transport calculations}
Current-induced spin accumulation was evaluated as a post-processing step using the tight-binding (TB) approach implemented in the \textsc{PAOFLOW} code.\cite{paoflow1, paoflow2} We have started with the \textit{ab initio} wavefunctions, projecting them onto the pseudoatomic orbitals in order to construct accurate tight-binding (PAO) Hamiltonians.\cite{agapito1, agapito2} We further interpolated these Hamiltonians onto ultra-dense $k$-points meshes of $140 \times 140 \times 110$ for Te and $80 \times 80 \times 60$ for TaSi$_{2}$, and we directly evaluated the quantities that are required to compute spin accumulation from Eq.(\ref{tensor}). In particular, the group velocities $\mathbf{v_k}$ are calculated as Hamiltonian's gradients $(1/\hbar){dH/d\mathbf{k}}$, and the spin polarization of each eigenstate $\psi(\mathbf{k})$ is automatically taken as the expectation value of the spin operator $\langle\psi(\mathbf{k})|\mathbf{S}|\psi(\mathbf{k})\rangle$.\cite{paoflow2} For an arbitrary energy $E$, the derivative of the Fermi distribution at zero temperature is equal to $\delta (E_\mathbf{k}-E)$ which we approximated with a Gaussian function. For  $T>0$, we explicitly calculated the derivative of the Fermi-Dirac distribution. We note that the adaptive smearing method could not be applied in this case, and a larger $k$-grid of $140 \times 140 \times 100$ had to be used to converge the calculation of TaSi$_2$ (Fig.~4c). The influence of the temperature on the electronic structures was not taken into account.


\section*{Code availability}
The PAOFLOW code can be downloaded from http://aflowlib.org/src/paoflow/.

\begin{acknowledgments}
We are grateful to Bart van Wees, Xu Yang and Caspar van der Wal for the insightful discussions that inspired us to initiate this project. J.S. acknowledges the Rosalind Franklin Fellowship from the University of Groningen. The calculations were carried out on the Dutch national e-infrastructure with the support of SURF Cooperative (EINF-2070), on the Peregrine high performance computing cluster of the University of Groningen and in the Texas Advanced Computing Center at the University of Texas, Austin.
\end{acknowledgments}

\section*{Author contribution}
J.S., A.R. and K.T performed the DFT calculations. A.R. conceived and analyzed model Hamiltonians. J.S, F.C, M.B.N, A.R. and A.J. implemented the method. A.R. and J.S. compiled the figures and wrote the manuscript. J.S. proposed and supervised the project. All authors contributed to the analysis and discussion of the results.




\end{document}